# Formal Verification of Solidity contracts in Event-B

Jian ZHU[1], Kai HU[1], Mamoun FILALI[3],
Jean-Paul BODEVEIX[3], and Jean-Pierre Talpin[2]

[1]State Key Laboratory of Software Development Environment, Beihng University
Beijing, China
[2]INRIA,Campus de Beaulieu
Rennes,France
[3]IRIT-CNRS,Univerisite de Toulouse
Toulouse, France
{zhujian, hukai}@buaa.edu.cn, { mamoun.filali, jean-paul.bodeveix }@irit.fr,
jean-pierre.talpin@inria.fr

**Abstract.** Smart contracts are the artifact of the blockchain that provide immutable and verifiable specifications of physical transactions. Solidity is a domain-specific programming language with the purpose of defining smart contracts. It aims at reducing the transaction costs occasioned by the execution of contracts on the distributed ledgers such as the Ethereum. However, Solidity contracts need to adhere safety and security requirements that require formal verification and certification. This paper proposes a method to meet such requirements by translating Solidity contracts to Event-B models, supporting certification. To that purpose, we define a restrained Solidity subset and a transfer function which translates Solidity contracts to Event-B models. Then we take advantage of Event-B method capabilities to refine models at different levels of abstraction to verify Solidity contracts' properties. And we can verify the generated proof obligations of the Event-B model with the help of the Rodin platform.

## 1 Introduction

The blockchain [1] is a distributed storage technology that provides safe and immutable storage of multi-party transactions by using fault tolerance and encryptions of data block chains linking the successive transactions. Once recorded in the chain, the data of a given block cannot be altered without altering all subsequent blocks, requiring consensus with the majority of parties in the network. Ethereum is a blockchain-based distributed computing platform which provides a decentralized virtual machine for executing scripts, such as Solidity contracts, using an internationally distributed network of public nodes [2].

A smart contract is the specification of contract terms using an algorithm that execute automatically when its pre-conditions are met (e.g. online user agreements).

In recent modern society, smart contracts transfer manually signed contracts on virtual networks and store them in a secure database or distributed ledger.

Solidity [3] is one of the popular specification languages to write smart contracts for the blockchain. It is a high-level object-oriented programming language for writing smart contracts dedicated to the Ethereum Virtual Machine. Solidity is statically typed, supports inheritance, libraries, and user-defined types, among other features. While it is usually quite easy to build a well-typed program that works as expected, it is much harder to verify that a smart contract cannot be used in an unanticipated or aggrieving manner.

Safety and security are hence essential properties for Solidity contracts, as valuable assets can be transferred through the transactions they automate. The potential security risks a Solidity contract exposes are to exploit vulnerabilities in either of the blockchain infrastructure or an underspecified contract algorithm (intentionally or not) to bias transactions and cause losses to other parties. The theft of funds in the DAO project of Ethereum [4] is one of the most significant example of security issues of smart contracts in the blockchain, which resulted in the loss of 60 million dollars from the account. Moreover, in April 2018, the BEC was exposed for security vulnerabilities and was attacked by hackers in the ERC-20 smart contract of Ethereum, causing an immense price crash which, if repeated, would lead the public to eventually distrust the technology altogether.

To try mitigating the reoccurrence of such events, developers have favored open-sourcing infrastructures for the block chains and making individual smart contracts and their execution public, as well as strengthening the security of contract implementations, using formal methods for verification. Formal method is a method of describing and reasoning computer system by mathematical method, and using formal reasoning for the development and analyses of a formal model for the given system can increase opportunities for finding errors in the system under test. The cost variation of applying the different formal methods against the programming model is shown in Fig. 1[5]. In this paper, we use Event-B method [5] which is a theorem proving method with a highest cost; therefore, it is suitably applied to our situations where more risk of funds security involved.

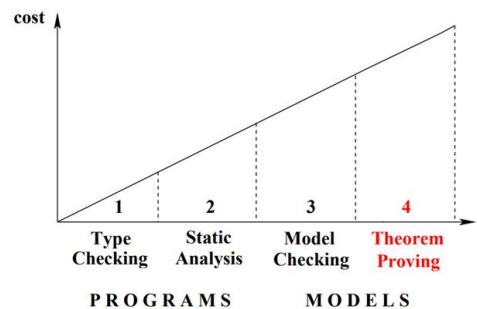

**Fig. 1** Formal verification methods versus cost

Event-B is a formal verification environment based on a typed set theory to specify and implement algorithms and systems as discrete transition systems [5]. It consists of two components: contexts and machines.

- A context is made of constants linked to some properties that define axioms, and sets that define data types.
- A machine has variables associated to invariants, and events. An event consists of a guard and an action. The guard denotes the enabling condition of the event, and the action denotes the way the event modifies the state.

As shown in Fig. 2, the dynamics of the model is "Events", which contains guards and actions. And they must satisfy the properties of constants and the invariants of variables, which belong to the static part.

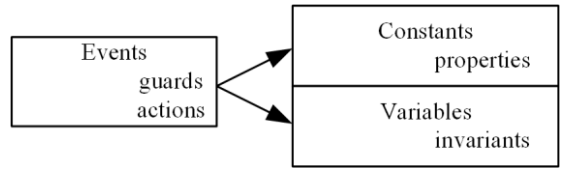

**Fig. 2** Structure of the Event-B model

The work presented in this paper is motivated by the need to build tools and techniques to improve security of smart contract by formal verification. We verify certain functionality and security requirements for the Solidity contract by translating it to the corresponding Event-B model with an appropriate transfer function, and then we simulate the model to validate the contracts' correct functionality and verify properties related to the Solidity contract using Event-B method. For translating Solidity contracts to Event-B [6] first-order logic, the Solidity language semantics should be well defined and understood to be semantically translated to Event-B language [7]. Our translation allows the validation of the functionality and verification of the properties related to the Solidity contract while preserving the semantics of the Solidity contract in the Event-B model.

Once verified, the contract can then be safely deployed on a public network. If one generated proof obligation cannot be verified, it will help us locate the problem which originates in the failure to prove the required invariants.

## 2    Related work

Solidity contracts have been widely used on the Ethereum platform for many companies to launch projects. Writing secure and safe smart contracts can be extremely difficult due to various business logics, as well as platform vulnerabilities and limitations. Formal methods have been advocated to mitigate these vulnerabilities. This paper proposes Event-B method to model and verify the Solidity contracts to improve their security, which complements some other articles on the same topic. In parallel with this work, there is some related formalization research on Solidity contracts and blockchain platform as follows.

The SMT-based formal verification module within the compiler of Solidity contracts has been described by Leonardo Alt [11]. SMT solver is a formal verification tool that checks the module automatically and it has been used to prove that the contract code satisfies the specification given by "require/assert" statements.

It considers "assert" statements as assumptions and tries to prove that the conditions inside "assert" statements are always true. If an assertion fails, the SMT solver generates a counterexample for the user, indicating the possible violation scenario.

Yuepeng Wang et al. [12] described a formal verification tool VERISOL towards smart contracts verification and bug-finding. However, it cannot apply to a general smart contract but a given one which implements the underlying workflow policy expressed as finite state machines. What's more, model checking method will confront state explosion when given system gets very complicated. Our approach to verifying properties of general Solidity contracts is based on first-order theorem proving, which is not limited to any particular specification.

Bhargavan K [13] used early a theorem proving technique for checking correctness of Solidity contracts by translation to F*, which is an interesting and pioneering research but is very complex like monads for side effects, dependent types and interaction with SMT provers. Another imperative language Dafny has the similar mechanism of proof, which is simpler to use. We have done a similar work with the simpler verification method in Event-B, which also supports a refinement of the model to verify more precise properties.

Grishchenko et al. formalized the EVM in F* [14] and they have run EVM tests to show the reliability of their model. But they haven't proved properties of any concrete contracts. Instead they consider classes of bugs in smart contracts and try to define general properties that prevent these. We have proposed a general verification framework and take a concrete Solidity contract honeypot as an example at the same time.

Ton Chanh Le et al. [15] determined the input conditions for which a smart contract terminates (or does not terminate) by proving conditional termination and non-termination statically. This is done by making sure that both, current state of the smart contract and the contract's input satisfy the termination condition to run on a proof carrying blockchain before the actual execution of the contract. They focus on the termination of smart contracts while we propose a general verification approach to the functionality of smart contracts.

Jakob Botsch Nielsen et al. [16] have modeled a vulnerable contract, faithful to the real DAO, and showed that by modeling it by Coq in a natural way one could have caught this vulnerability. They have also implemented an improved version of the congress and showed that it does not have this vulnerability.In general, it's an interesting and available research to find bugs, and we have proposed a similar method by modeling smart contracts in Event-B but we emphasize on verifying the different levels of properties of smart contracts while they focused on the transactions and the contract calls.

## 3  From Solidity contracts to Event-B models

Based on the previous introduction of definitions, we establish a semantical map between Solidity contracts and Event-B language. Fig. 3 outlines the framework to analyze and formally verify Solidity contracts using Event-B method. Once we have the Solidity contact translated to Event-B model, its correctness is established by proof obligations for the invariants, which should be preserved by each event, including the initialization event. We first analyze the syntax of Solidity contracts to define an appropriate transfer function, which takes the solidity contract as input, and outputs the corresponding abstract Event-B model. The most important part of the smart contract is its properties and constraints. Some of them are explicit in the form of a statement like "require", which we can translate directly to specific Event-B guards according to the translation rules. As we have mentioned before, Event-B guards are used to define preconditions that should hold before the event can be executed. Others are implicit like the requirement that the total balance of funds be constant during a safe remote purchase transaction, which can be concluded manually by refining our abstract model. In our refined model, we define similar events to model the updates on the abstract one. And the correctness of the gluing invariant over the two models can keep the correctness of the Solidity contract, which is modeled by the invariants with respect to the Event-B model.

Each main component of Solidity contracts like type declaration, attributes, constructor and functions are modeled in Event-B. And properties related to the correct operation of the Solidity contract are modeled as Event-B invariants. Rodin platform [8] is used to check these invariants and validate the correct functionality of events using simulation as shown in Fig. 3. Once the result of verification is false, we can find the location of errors and go to the modification part to modify our smart contracts; if everything is ok, we can obtain a certificate for our smart contracts and apply them. In the following section, we define a transfer function based on Solidity subset and explain the translation rules with a concrete Solidity contract.

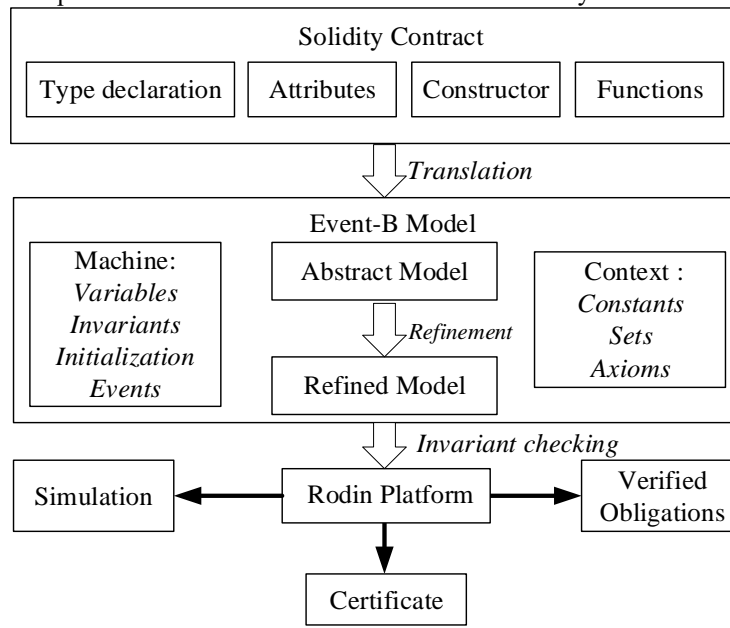

**Fig. 3** Translation and verification of Solidity contracts in Event-B

## 3.1 Solidity Subset Grammar

Solidity is a contract-oriented, high-level language whose syntax is similar to that of JavaScript and it is designed to target the Ethereum Virtual Machine. Here we consider a restricted subset of solidity shown in Fig. 4, which contains the main functionality of this language. And we predefine three types of statements in solidity:
- Arithmetic expressions AExp (elements are denoted α, $α_0$, $α_1$…)
- Boolean expressions BExp (elements are denoted b, $b_0$, $b_1$…)
- Commands Com (elements are denoted c, $c_0$, $c_1$…)

Let n, $n_0$, $n_1$…denote integers, x, $x_0$, $x_1$…denote a countable set of variables, t, $t_0$, $t_1$…denote a countable set of variables addr denote type '**address**' in solidity. '**Exp**' contains <**AExp**> and <**BExp**>.

<method>   m::='function'(@identifier)? '()'(<qualifier>)*
           '{x:=Exp? ';')+)?
           (<Com> ';')* '}'
<type>     t::='integer'| 'address'| 'bool'| 'mapping'$(t_1 \Rightarrow t_2)$
<AExp>     a::=n|x|byte|addr| $a_0 \oplus a_1$
<BExp>     b::=true|false|$a_0 \odot a_1$|$b_0 \oslash b_1$|$\neg b$
<Com>      c::=x:= $a$| $c_0; c_1$ |if $(b)$ $c_1$ else $c_2$| require $b$ do $c_1$
<qualifier> q::='private'| 'public'| 'internal'
           | 'external'| 'returns (' <type>(@identifier)? ')'

$\oplus ::= +| * | - | /$
$\odot ::= \leq | \geq | == |! =$
$\oslash ::= \vee | \wedge$

**Fig. 4** Syntax of the Solidity subset written in BNF notation

Notably, the subset of solidity defined here does not include loops, which needs to be translated to a very complex structure in Event-B model. Although it does not affect our case study but we try to solve it in the future. The three main types of declarations within a contract are type declarations, property declarations, and methods. Type declarations mainly consist of integers, strings, address, structs, enums and mappings. Property declarations are reflected in keywords like "public" and "payable". Any user or contract can call or access the variables and functions decorated by "public". Payable is a special method for receiving Ethereum; for example, when we invoke any function in Solidity contracts with "Payable", we need to transfer tokens (i.e. msg.value) not lower than specified values before the body of the function executes. State variables (variables declared outside the function) default to the "storage" form and are permanently written to the blockchain; variables declared inside the function default to "memory" type, and they disappear after the function call ends. Methods are compiled in Ethereum virtual machine into a single function that runs when a transaction is sent to the contract's address.

### 3.2 Smart contract honeypot Example: Gift_1_eth

As an example, we consider a concrete smart contract honeypot, whose source code can be found on the GitHub website [9]. Honeypot contracts are such contracts that hold ethers but pretend to do so in an insecure manner, fooling hackers into thinking that they can steal the ethers from those contracts. Its source code is available for everyone to analyze and modify. As a lot of variation derived from Gift_1_eth exist in blockchains, some have the same source code, while others make some minor changes. In this case study, we present one of them in Table 1 and translate it to Event-B model according the following rules. Then we use Event-B method to analyze and verify the properties related to the Solidity contract. There are three main functions to ensure the transaction go smoothly.

- SetPass (): When the sender's transaction value is bigger than 1 ether and the variable passHasBeenSet is set to false, it can set the new password (i.e. hashPass).
- GetGift (): When the password entered is equal to the set value (i.e. hashPass), the sender can take all ethers in the contract.
- PassHasBeenSet (): if the password entered equals to the set value (i.e. hashPass), then the variable passHasBeenSet will be set to true.

**Table 1** Solidity contract source code

```
1   pragma solidity ^0.4.17;
2   contract Gift_1_ETH
3   {
4       bool passHasBeenSet = false;
5       function()payable{}
6       function GetHash(bytes pass)  constant  returns
7   (bytes32) {return sha3(pass);}
8       bytes32 public hashPass;
9       function SetPass(bytes32 hash)
10      payable
11      {
12          if(!passHasBeenSet&&(msg.value >= 1 ether))
13          {
14              hashPass = hash;
15          }
16      }
17      function GetGift(bytes pass) returns (bytes32)
18      {
19  
20          if( hashPass == sha3(pass))
21          {
22              msg.sender.transfer(this.balance);
23          }
24          return sha3(pass);
25      }
26      function PassHasBeenSet(bytes32 hash)
27      {
28          if(hash==hashPass)
29          {
                passHasBeenSet=true;
```

|  |  |  |
|---|---|---|
| 30 |  | } |
| 31 |  | } |
| 32 | } |  |
| 33 |  |  |

The typical Solidity contract shown in Table 1 is made up of four components, such as type declaration, attributes, constructor and functions. In particular, it consists of a single-entry point that decides on which method code to invoke upon the incoming transaction (such as a received message **msg**). Its methods have access to global variables that contain information about the contract (such as the current balance of the contract in **this.balance**), the transaction used to invoke the contract's method (such as the source address in **msg.sender**, and the amount of Ether received in **msg.value**). It supports several methods of transferring ethers between the contracts (like **msg. sender. transfer(amount)**).

### 3.3 Translation from Solidity to Event-B

As shown in in Fig. 5, we define a transfer function T who translates our solidity syntax to Event-B while preserving the semantics. Here the input is the syntax elements of Solidity subset defined in Fig. 4 like a, b, c, t and the function of the Solidity subset. The transfer function T outputs the corresponding syntax elements of Event-B. We define such a general translation mapping for the solidity contract:
- Type declaration is translated to sets and axioms in Event-B.
- Attributes are translated to variables and invariants in Event-B.
- Constructor is translated to initialization in Event-B.
- Functions are translated to events, constants and axioms in Event-B.

$T[m] ::= $ 'function'(@identifier)?'()'(<qualifier>)*
  '{(x:=Exp?,')+)?
  (<Com>';')*'}' $\Rightarrow$
  'event' (@identifier) where (guard) then (x:=Exp)

$T[t] ::= $ case t of
  | integer $\Rightarrow$ integer
  | address $\Rightarrow addr_{set}$
  | bool $\Rightarrow$ bool
  | mapping $\Rightarrow t_1 \rightarrow t_2$

$T[b] ::= $ case b of
  | true $\Rightarrow$ true
  | false $\Rightarrow$ false
  | $a_0 \odot a_1 \Rightarrow A_0 \odot A_1$
  | $b_0 \oslash b_1 \Rightarrow B_0 \oslash B_1$

$T[a] ::= $ case a of
  | byte $\Rightarrow$ integer
  | n $\Rightarrow$ N
  | x $\Rightarrow$ X
  | $a_0 \oplus a_1 \Rightarrow A_0 \oplus A_1$

$T[c] ::= $ case c of
  | x:=a $\Rightarrow$ x:=e
  | $c_0; c_1 \Rightarrow$ act1;act2
  | if (b) $c_1$ else $c_2 \Rightarrow$ {TRUE$\mapsto c_1$,FALSE$\mapsto c_2$}(bool(b))
  | require b do c $\Rightarrow$ where g then act

**Fig. 5** Transfer function of the Solidity subset

Then we structurally translate Solidity to Event-B by giving translation rules for each construct separately and in detail.

*1)* Contracts are translated to Event-B project, which contains Event-B machines and Event-B context.

*2)* Type declaration is translated to sets and axioms in a context. For example, if Solidity contracts use a data type like unsigned integer which is the same in Event-B language, then it's easy to translate it. If not, we construct a corresponding set to represent it. For example, the address type is unique in Solidity contracts, and thus we abstract it as a set called "ADDRESS" in Event-B. And we defined a variable "address_tem" as a current set recoding all the addresses who call this contract. As shown in Fig. 6.

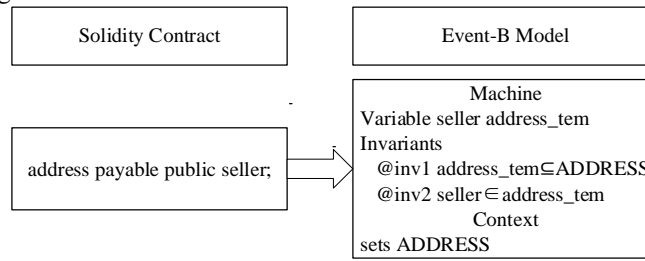

**Fig. 6** Translation of the type declaration of "seller"

*3)* Attributes are replaced by variables and invariants, and the invariants here are used to specify the type of variables, or else we do not know which type of the variable is. All variables have defaulted to property "public" and they need initialization, as shown in Fig. 6.

*4)* The constructor is replaced by initialization. The input parameters are replaced by constants defined in the context. What's more, initialization in Event-B will contain initialization of all variables which is different with the constructor in Solidity contracts. Here we take the Gift_1_eth contract as example. Its constructor only contains a "payable" keyword which means all other address invoke this contract need to pay specified ethers. It doesn't affect our Event-B model because the value of the balance initialized is arbitrary, and thus the translated initialization event contains only initialization of four variables. As shown in Fig. 7.

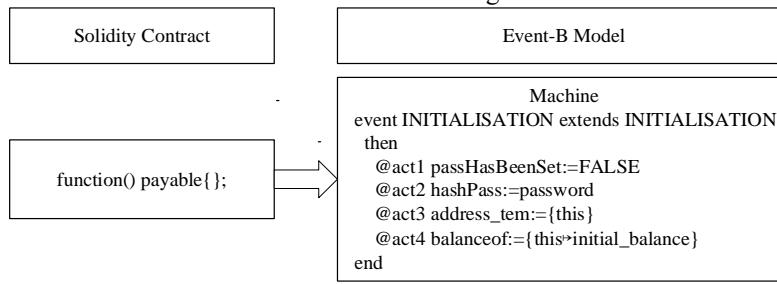

**Fig. 7** Translation of the constructor

*5)* Functions are translated into events. In this part, require/assert statements are very important as a premise of successfully calling the function, and we translate them to Event-B guards in the events. If statement is translated to our specified function in Event-B as already shown in the transfer function T. If some functions are decorated with "payable", we will create a variable called "balanceof" in Event-B machine, which is a map from addresses to their balances. At the same time, we create two necessary parameters called "msg_sender" and "msg_value". Besides, we need to define a particular event called "NewAccount", and it represents the new address which invokes this contract. If statement is translated to a specified function structure mentioned in Fig. 5. Here we take the function PassHasBeenSet as example, where we translate the data type byte32 to the integer in Event-B because it doesn't matter the functionality and it's simple to show. As shown in Fig. 8.

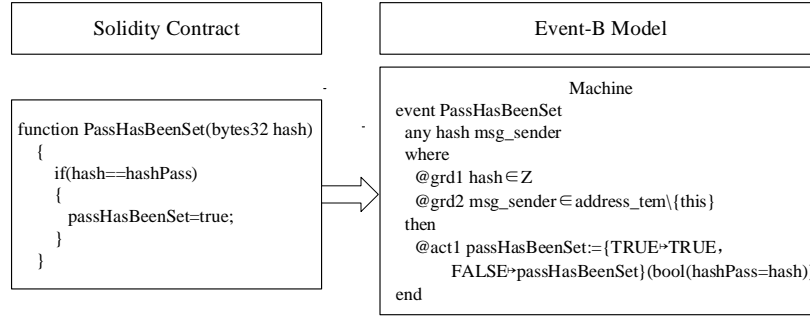

**Fig. 8** Translation of the function

Based on the previous translation rules, we have finished a corresponding translation of the Solidity contract with an Event-B abstract machine named Gfit_ether_m1 and the Event-B context named Gift_ether_c. We have put the source code of the whole Event-B model on the GitHub website [10], where we also have put another verified Event-B model translated from a famous and typical Solidity contract named SafeRemotePurchase. Next, we present the remaining important parts of the Event-B abstract model expect for what we have already shown in previous translation rules.

In Table 2, we have modeled a data entity, whose first part defines an abstract set "ADDRESS", which we will use as the source of unique identifiers for different addresses. In the second part, "this" represents the address of the contract itself, which is necessary for every translated model. "password" is the initial value of the variable "hashPass" in the Event-B machine and "initial_balance" is the initial balance of the contract address "this". Besides, "TRANSFER_VALUE" is the default of the minimum transaction value when calling a function decorated with a keyword "payable". For convenience, we use the integer as the common data type for the last three constants. Although their data types will be more precise and a little different in Solidity contracts, it does not affect our modelling of the contract functionality. When saving an Event-B component, Rodin starts type checker to ensure that types are correctly used.

**Table 2** Event-B's context model

```
context Gift_ether_c
sets ADDRESS
constants this password initial_balance TRANSFER_VALUE
axioms
  @axm1 this∈ADDRESS
  @axm2 password∈Z
  @axm3 initial_balance∈N1
```

```
    @axm4 TRANSFER_VALUE∈N1
  end
```

In the event SetPass as shown in Table 3, we define three parameters and five guards, where @grd4 requires that the value transferred cannot exceed the balance and @grd5 requires that the value transferred should be higher than the default. These two guards keep the properties of the Solidity contract. Then we define two actions: @act1 is obtained by using the translation rules applied to the if statement, and @act2 is obtained by using the translation rules applied to the keyword "payable".

**Table 3** Event "SetPass" of the Event-B model

```
event SetPass
  any hash msg_sender msg_value
  where
    @grd1 hash∈Z
    @grd2 msg_sender∈address_tem\ {this}
    @grd3 msg_value∈N1
    @grd4 msg_value≤balanceof(msg_sender)
    @grd5 msg_value≥TRANSFER_VALUE
  then
    @act1 hashPass:={TRUE↦hash,FALSE↦hashPass}
          (bool(passHasBeenSet=FALSE
          ∧msg_value≥TRANSFER_VALUE))
    @act2 balanceof:=balanceof<+{this↦
           balanceof(this)+msg_value
          ,msg_sender ↦balanceof(msg_sender)-msg_value}
end
```

## 4  Simulation and verification

After finishing the translation work, we can obtain an abstract Event-B model, which should perform the correct behavior as the Solidity contract does. We use the simulation function of a Rodin's plugin called ProB. It serves to validate that the model's behaviors are consistent with the design, which can easily help us to simulate the Solidity contract. As shown in Fig. 9, we have simulated the abstract model in Rodin by assigning different values to the parameters and result shows that our model performs the correct behaviors as we expected, it's a process of interacting with the tool.

| Name | Value | Previous value |
|---|---|---|
| ∨ Gift_ether_c | | |
| TRANSFER_VALUE | 1 | 1 |
| initial_balance | 0 | 0 |
| password | 0 | 0 |
| ∨ ★ Gift_ether_m1 | | |
| address_tem | {this,ADDRESS2} | {this,ADDRESS2} |
| ★ balanceof | {(this↦1),(ADDRE... | {(this↦0),(ADDRE... |
| hashPass | 0 | 0 |
| passHasBeenSet | TRUE | TRUE |
| ∨ Formulas | | |
| > sets | | |
| > invariants | T | T |
| > axioms | T | T |
| > event guards | | |
| invariants ok | | no event errors detected |

**Fig. 9** Simulation of the Event-B model

Besides, each property generates a number of proofs obligations. These proof obligations are proven one by one, some are automatically discharged using the proof tools like SMT provers, and some need to be proven interactively by providing certain rewrite rules to simplify the obligation. In the following, we state two important properties for guaranteeing the security of the Solidity contract. The first one is defined for the abstract machine Gift_ether_m1, and it states that the balance of each account should be strictly positive or zero.

**Prop. 1** balanceof ∈ address_tem → $\mathbb{N}$

As shown in Fig. 10, it is defined as invariants in Event-B model, the tool generates proof obligations which are successfully discharged using Event-B proof control. (a completed proof is indicated by a green mark).

```
📂 Gift_ether
  > 🅖 Gift_ether_c
  ∨ Ⓜ Gift_ether_m1
      > ● Variables
      > ✦ Invariants
      > ✹ Events
      ∨ ✅ Proof Obligations
            🅰 INITIALISATION/inv4/INV
            🅰 INITIALISATION/inv5/INV
            🅰 NewAccount/inv4/INV
            🅰 NewAccount/inv5/INV
            🅰 SetPass/grd4/WD
            🅰 SetPass/inv4/INV
            🅰 SetPass/act1/WD
            🅰 SetPass/act2/WD
            ✅ GetGift/inv4/INV
            🅰 GetGift/act1/WD
            🅰 PassHasBeenSet/act1/WD
```

**Fig. 10** Verified proofs obligations of the Event-B model

The second one is that the balance of the sender should not be changed while the variable passHasBeenSet is true (we don't consider the cost of the mining), which means that we shouldn't pay for the function if it can't be used by us.

**Prop. 2** $\forall i \cdot i \in address\_tem$, passHasBeenSet=FALSE $\Rightarrow$

   Balanceof (i) [Before SetPass] =balaceof (i) [After SetPass]

This property prevents the function from stealing funds from our account, which cannot be expressed in the abstract model, and we need to construct a refined model Gift_ether_m2 to make the event SetPass precise and verify this property. As shown in Table 4, we emphasize the condition that passHasBeenSet equals to true in @grd6, and the @act2 states that the balance should not be changed.

**Table 4** Refined event "SetPass" of the Event-B model

```
event SetPass refines SetPass
  any hash msg_sender msg_value
  where
    @grd1 hash∈Z
    @grd2 msg_sender∈address_tem\{this}
    @grd3 msg_value∈N1
    @grd4 msg_value⩽balanceof(msg_sender)
    @grd5 msg_value⩾TRANSFER_VALUE
    @grd6 passHasBeenSet=TRUE
  then
    @act1 hashPass≔ hashPass
    @act2 balanceof≔ balanceof
end
```

As shown in Fig. 11, the result shows that there is one proof cannot be completed, it points out that our Solidity contract has a logical vulnerability in SetPass function because it cannot satisfy our second property. In fact, before the contract is attacked, the contract creator sets a password that only he knows and set the variable passHasBeenSet to true, so that only the contract creator can take out the ether in the contract. Others who try to call the SetPass function will loss at least the default ether, which should not happen! And that's the point of this contract honeypot. Our Event-B model can be further refined in order to include more properties about the Solidity contract, while preserving the correctness of the invariants.

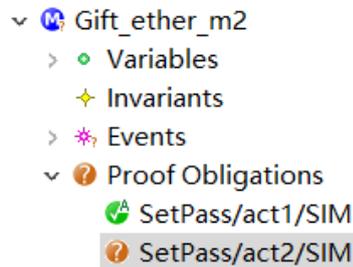

**Fig. 11** Pop up error presented by the Rodin platform

## 5   Conclusion

In this paper, we propose an approach for translating Solidity contracts to Event-B, making a contribution to enhance security of Solidity contracts by verifying its properties via formal verification. For this purpose, we have defined the transfer function from Solidity subset languages to the Event-B model while preserving the semantics. To better demonstrate the practicability of our approach, we have translated a Solidity contract honeypot to the Event-B model according to the translation rules. We use the ProB tool to simulate our abstract model to validate that it performs the expected behaviors, and some basic properties verification and type checking are automatically done by Rodin. Finally, we successfully found the logical vulnerability by formally refinement in Event-B. We found it is of great importance to apply this approach to other Solidity contracts involved huge funds transactions.

The approach we present here is based on first-order theorem proving, it provides functional verification of the Solidity contract at different levels of abstraction. The rich expressive language of the first-order logic allows us to verify complex properties of Solidity contracts. As future work, we will extend current Solidity contracts subset to more types and features, and establish a more refined description of Solidity contracts which will allow the verification of more detailed properties.

We also plan to realize a translator tool for more efficient and more practical translation from Solidity contracts to Event-B models. The algorithm should follow the translation rules defined in our approach. It can help smart contracts developers who are not familiar with formal methods to improve the security of smart contracts.